# A dispersal-recolonisation 3D biofilm *in vitro* model based on co-assembled peptide amphiphiles and clinical wound fluid

Zhiquan Yu[1,2,#], Chenjia Zhao[3,#], Lingyun Xiong[1,2], Shanshan Su[1,2], Dawen Yu[1,2], Shilu Zhang[1,2], Yubin Ke[4], Hua Yang[4], Guo Zhang[1,2], Jiaming Sun[1,2], Nengqiang Guo[1,2,*], Yuanhao Wu[1,2,*]

[1]Plastic and Reconstructive Surgery Department, Wuhan Union Hospital, Tongji Medical College, Huazhong University of Science and Technology, Wuhan, China.

[2]Plastic and Reconstructive Surgery Research Institute, Wuhan Union Hospital, Tongji Medical College, Huazhong University of Science and Technology, Wuhan, China.

[3]China Academy of Aerospace System and Innovation, Beijing, 100088, China

[4]Spallation Neutron Source Science Center, Dalang, Dongguan, China.

#These authors contributed equally to this work.

*To whom correspondences should be addressed.

E-mails: nengqiangguo@hust.edu.cn; yuanhaowu@hust.edu.cn

**Abstract**

Chronic wound infections are sustained by dynamic 3D biofilm cycles involving maturation, dispersal, and recolonisation, yet existing *in vitro* models fail to reproduce these temporal and structural complexities. Here, we report a strategy that co-assembles a designed protease-inhibitory peptide amphiphile (PA-GF) with patient-derived wound fluid (WF) to reconstruct the complete biofilm life cycle *in vitro*. The PA-GF sequence incorporates an HWGF motif capable of binding and inhibiting matrix metalloproteinase-9 (MMP-9), thereby preserving the integrity of recolonised biofilms under proteolytic stress. Co-assembling with WF generated a living material that faithfully mimicked the biochemical and mechanical microenvironment of chronic wounds, supporting the formation of stable 3D biofilms capable of dispersal and recolonisation. Furthermore, we established a controllable polymicrobial infection model and validated its translational relevance through antibiotic susceptibility profiling and spatial microbiological analyses. Notably, the antibiotic response patterns of the PA/WF-derived biofilms closely mirrored those observed in a rat wound infection *in vivo* model. Collectively, our findings demonstrate that co-assembling living materials can recapitulate the nutritional composition, 3D architecture, and recolonisation dynamics of *in vivo* infectious biofilms, offering a physiologically relevant and customisable platform for investigating chronic wound infections and accelerating anti-biofilm drug discovery.

**Introduction**

Chronic wounds, such as diabetic foot ulcers and pressure sores, impose a substantial global socioeconomic burden[1,2]. A major pathological hallmark underpinning their persistence is the formation of three-dimensional (3D) biofilms by pathogenic bacteria on the wound surface[3,4]. Biofilms are structured microbial communities embedded in a self-produced extracellular polymeric substance (EPS) matrix that adheres to tissue surfaces[5,6]. This matrix acts as a physical and chemical barrier, limiting antibiotic penetration and shielding bacteria from host immune responses, thus complicating infection clearance[7–9]. To devise more effective therapeutic strategies, it is essential to faithfully recapitulate biofilm formation, architecture, and drug responsiveness under *in vivo* conditions[10]. Alternatively, *in vitro* biofilm models have emerged as powerful platforms for anti-biofilm drug screening and mechanistic studies, offering advantages such as experimental controllability, reproducibility, and ethical feasibility[11–13]. Advances in materials science and bioengineering have significantly enhanced the sophistication of these models[14,15]. Ning *et al.* employed 3D printing to fabricate alginate hydrogel scaffolds that replicate the spatial complexity of biofilms[16], while Lubbock *et al.* increased physiological relevance by incorporating whole blood or serum to better simulate the host immune microenvironment[17,18]. These developments have greatly facilitated early-stage evaluation of anti-biofilm interventions. However, chronic wound biofilm infections are inherently dynamic processes, but existing models inadequately reflect the temporal evolution of biofilms beyond their static structural and compositional complexity[19,20].

The dynamic property of *in vivo* biofilm is characterised by "maturation" to "dispersal" and subsequent "recolonisation" process[21–24]. During this cycle, bacteria released from mature biofilms can actively recruit and incorporate host matrix proteins such as fibrin, fibronectin, and collagen into newly forming biofilms[25,26]. Dispersal-recolonisation plays pivotal role in infection dissemination, sepsis onset, and chronic disease progression, as well as in fostering therapeutic resistance[24,27,28]. Nonetheless, current *in vitro* models primarily focus on the initial attachment and maturation phases, with limited capacity to emulate dispersal–recolonisation dynamics[10]. In a notable attempt, Song *et al.* utilised a PDMS-based microfluidic platform to model recolonisation; however, due to the inadequate affinity between PDMS and biofilm components, only two-dimensional surface-attached microcolonies could be formed[29–31]. Thus, the key challenge in recapitulating the dynamic biofilm life cycle lies in the development of living materials that support robust bacterial colonization. In addition, these materials must respond adaptively to microenvironmental fluctuations that occur throughout the different stages of biofilm development[32,33].

Supramolecular co-assembly offers a promising strategy for modelling this dynamic process. By rationally designing co-assembly monomers, it is possible to enhance both structural and chemical complexity, resulting in biomimetic materials with hierarchical architectures[34], self-healing properties[35], and environmental responsiveness[36]. Relying on non-covalent interactions, supramolecular systems inherently exhibit reversibility and adaptability, enabling real-time responsiveness to microenvironmental cues such as pH[37], ionic strength[38], and temperature[39], which dynamically fluctuate during biofilm development. Our previous studies have demonstrated that peptide amphiphile (PA) can co-assemble with artificial sputum medium (ASM) supporting 3D biofilm formation in modelling respiratory tract infections, thereby establishing a foundation for further exploration[40]. Nonetheless, two critical challenges must be addressed to recreate a biofilm infection that more closely mimics *in vivo* chronic wound conditions. First, the inability to reconstruct the dynamic process of 3D biofilm dispersal and recolonisation; Second, it is necessary to involve the direct integration of real host endogenous molecules present at the infection site, enabling a wound-specific customization of nutrient composition[10].

To address these challenges, pioneer studies of Mata group have demonstrated the feasibility of leveraging blood-derived biomolecules in conjunction with peptide amphiphiles (PAs) to construct personalized regenerative materials[41]. Inspired by this, patient-derived wound fluid (WF) is a potential co-assembly component[42,43]. The microenvironment of chronic wound-associated biofilm growth is known to vary substantially depending on host-specific factors, anatomical location, and microbial species, making it difficult to establish a universal *in vitro* model that accurately recapitulates these complexities[44,45]. WF is not only readily accessible but also physiologically relevant, containing endogenous components including immune mediators, proteases, and metabolites that closely resemble the specific biochemical milieu of individual wound sites[46–48]. Specifically, we have observed that matrix metalloproteinase-9 (MMP-9), a protease typically upregulated in chronic wounds exhibits aberrant proteolytic activity *in vitro*, which is potential to compromise PA fibres integrity during the recolonisation phase and ultimately disrupts the macrostructure of recolonized biofilms[49–51].

Here, we report on an *in vitro* infection model that replicates the architectural and dynamic features of chronic wound biofilms. First, we have designed an innovative protease-responsive peptide amphiphile (PA-GF) by incorporating the HWGF motif, which enables MMP-9 binding and inhibition, helping preserve 3D biofilm structure under proteolytic stress. Second, we biofabricated the complete biofilm lifecycle, from maturation through dispersal to recolonisation, leading to the formation of stable and high-biomass recolonized biofilms. Third, we established a quantitative co-infection system using *Pseudomonas aeruginosa* (ATCC 27853) and *Staphylococcus aureus* (SH1000), allowing modulation of microbial dominance to mimic polymicrobial infections. We then standardised the platform for high-throughput antibiotic screening, which produced results consistent with those from conventional clinical assays. Finally, *in vivo* testing in a rat wound model revealed distinct microbial spatial patterns under antibiotic pressure, closely aligned with the *in vitro* phenotypes observed using the same treatments. Taken together, this platform offers a physiologically relevant and clinical translationally applicable tool for investigating dynamic 3D biofilm infections and accelerating anti-biofilm drug discovery.

**Results**

**1. Rationale of the system**

*Molecular design*

We selected positively charged peptide amphiphiles (PAs) incorporating the fibrinogen-derived AGD motif to promote rapid gelation upon contact with wound fluid (WF) or artificial wound fluid[40] (AWF) (Figure 1a). The WF was used to construct infected PA/WF hydrogels, simulating regions of *in situ* biofilm infection in chronic wounds. In contrast, AWF was formulated not only to reproduce the inorganic salt composition of chronic wound exudates but also supplemented with serum to mimic the nutritional environment. The PA/AWF hydrogel was therefore designed to represent biofilm-free regions, facilitating the establishment of recolonisation models to validate the system's ability to reproduce the dispersal–recolonisation dynamics observed *in vivo*. To enhance the stability of biofilm under protease-rich conditions, we designed the HWGF motif into the PA backbone as the GF sequence (Figure 1a), as this motif was known to bind matrix metalloproteinase-9[52] (MMP-9). Molecular docking of GF with MMP-9 (PDB ID: 1GKC) revealed multiple stabilising interactions, including hydrogen bonds, π–hydrogen contacts, and hydrophobic interactions, involving both catalytic residues (His401, His405, His411, Glu402) and adjacent binding site residues (Tyr423, Met422, Leu188, Leu418, Phe425, Pro421). These interactions anchor GF within the catalytic pocket, likely hindering substrate access and providing a mechanistic explanation for its inhibitory effect on MMP-9 (Figure 1b). The enzyme-inhibitory property of PA-GF was supposed to enhance the stability of the PA-supported biofilm throughout both its formation and recolonisation phases.

*Design of the dispersal-recolonisation 3D biofilm model for antibiotics screening kit*

In order to recapitulate the complete biofilm life cycle, we established a recolonisation assay using a mixed PA water solution of PA-GF and PA-AGD. When clinically extracted WF with pathogenic bacteria co-assembled with the PA solution and generated a hydrogel, bacteria were embedded within the hydrogel matrix. After 24 h of culture, a primary biofilm formed, which faithfully mimicked tissue infected by three-dimensional (3D) biofilm *in situ*. This step not only reproduced the initial colonization process observed in chronic wounds but also provided a physiologically relevant platform for subsequent infection modelling. The primary biofilm was then directly coupled to a clean serum-based PA/AWF hydrogel within a cell insert chamber.

The uncolonized hydrogel represented surrounding, uninfected tissue adjacent to an infected lesion. During prolonged culture, bacteria dispersed from the mature primary biofilm and colonized the clean hydrogel, generating a recolonized biofilm (Figure 1c). This process reproduced the clinically relevant transition from local infection to secondary colonization of neighbouring tissue, thereby validating that the model supports both dispersal and recolonisation phases of the 3D biofilm life cycle. We subsequently integrated this dynamic 3D biofilm model into a 24-transwell-based plate, positioning the recolonized biofilm in the upper chamber and adding antibiotics at graded concentrations to the lower chamber (Figure 1d). The resulting configuration constituted a high-throughput-compatible antibiotic screening platform, which allows systematic evaluation of antimicrobial efficacy under conditions that more closely mimic biofilm-associated chronic wound infections *in vivo*.

**2. Characterization of the PA/(A)WF co-assembly system**

*Supramolecular characterization of PA/(A)WF co-assembly*

In order to investigate the supramolecular features of PA/(A)WF system, we first demonstrated that PA-AGD and PA-GF carried surface charge of 56.27±5.05 mV and 40.33±3.59 mV, whereas both AWF and WF from chronic wounds carried negative surface charges as measured by electrophoretic light scattering (ELS) (Figure. 1a, Extended Data Fig. 1a). The particle size of PA-AGD increased substantially from ~187 nm to ~3540 nm upon co-assembling with WF, whereas that of PA-GF-AGD exhibited a more pronounced expansion, rising from ~163 nm to ~6,700 nm, as measured by dynamic light scattering (DLS, Figure. 2a), suggesting synergistic intermolecular interactions between PA-GF and PA-AGD. A comparable increase was observed for the particle sizes being co-assembled between PAs and AWF (Extended Data Fig. 1b), confirming the cooperative co-assembly behaviour of PA-GF and PA-AGD in both WF and AWF. To further evaluate the secondary structure alterations induced by co-assembly, circular dichroism (CD) analysis revealed that WF or AWF exhibited a characteristic *α*-helical signal, whereas PA-GF-AGD predominantly adopted a random coil conformation prior to co-assembly. Notably, the mixed co-assemblies showed a *β*-sheet-rich CD spectrum, indicating a conformational transition induced by co-assembly (Figure. 2b, Extended Data Fig. 1c). These findings confirmed that the interaction between the PA-GF-AGD and WF or AWF enhanced structural ordering and led to the formation of an organized supramolecular architecture.

*Nano- and Micro- scales Characterization of PA/(A)WF co-assembled structures*

To elucidate the interplay between PAs and (A)WF, we investigated their co-assembled hierarchical structures across nano- and micro-scales. First, the co-assembled structures exhibited classical and homogeneous nanofibrous networks revealed by scanning electron microscopy (SEM) (Figure. 2c, Extended Data Fig. 1d). Before co-assembling, PA-AGD self-assembled into rigid rod-like assemblies with a radius of gyration ($R$g) of ~232 Å, whereas PA-GF self-assembled into semi-flexible structures with an $R$g of ~302 Å, as determined by small-angle neutron scattering (SANS) (Figure. 2d). Upon co-assembling with WF, PA-AGD/WF aggregates displayed a distinct $q^{-1.7}$ scattering pattern in the intermediate-$q$ region, indicative of semi-flexible structures, with an increased $R$g of ~450 Å. By contrast, GF-AGD/WF assemblies exhibited a $q^{-2}$ dependence, consistent with a branched morphology, and an $R$g of ~350 Å (Figure. 2e). In both AGD/AWF and GF-AGD/AWF systems, scattering curves in the low-$q$ (Guinier) region showed a clear $q^{-3}$ dependence, implying the formation of solid-like 3D aggregates with smooth surfaces, whereas the high-$q$ (Porod) region revealed a $q^{-1}$ decay characteristic of internal rod-like structures (Extended Data Fig. 2a). The fitted $R$g values of these aggregates reached ~600 Å and ~500 Å, respectively. These nanoscale findings were corroborated by transmission electron microscopy (TEM), which revealed thick fibrillar structures corresponding with the SANS (Figure. 2f, Extended Data Fig. 2b). Collectively, these results demonstrate that PA-GF-AGD co-assemble with (A)WF into homogeneous hydrogels comprising hierarchically uniform nanofibres.

*Components in co-assembled hydrogels and the function of selectively inhibiting MMP-9*

To investigate the composition of PA/(A)WF co-assembled hydrogels, we first examined protein incorporation. Protein bands originating from WF were effectively retained within PA/WF hydrogels, indicating successful protein incorporation, as demonstrated by sodium dodecyl sulphate-polyacrylamide gel electrophoresis (SDS-PAGE, Figure. 2g). Protein level analysis further showed that 2% PA-AGD retained more WF proteins than 1% PA-AGD, and co-assemblies of 2% PA-AGD with 0.5% PA-GF retained even higher protein levels, increasing the concentration of PA-GF further enhanced protein retention, as quantified by band intensity analysis (Figure. 2h). Collectively, these results demonstrate that incorporation of PA-GF enhances WF protein retention within the hydrogel matrix. Furthermore, fluorometric assays (EnzChek™) demonstrated that PA-GF is a dose-dependent inhibitor of MMP-9, with activity detectable at 0.0001 wt%, whereby 0.01 wt% achieved more than 70% suppression of enzymatic function within 20 h. (Figure. 2i). Following this, we examined the biofilm-forming ability of gram-negative *Pseudomonas aeruginosa* ATCC 27853 (Gram$^-$, *P. aeruginosa)* and gram-positive *Staphylococcus aureus* SH1000 (Gram$^+$, *S. aureus*) on hydrogels through co-assembling of PA with either WF or AWF. These mature biofilms with dense microbial colonization closely integrated into the hydrogel matrix as revealed by SEM (Figure. 2j, Extended Data Fig. 2c). Subsequent antibiotic testing revealed that these biofilms displayed markedly enhanced antibiotic tolerance compared with their planktonic counterparts (Extended Data Fig. 3a). These findings confirm that PA-GF retains potent MMP-9 inhibitory activity and that PA/(A)WF co-assembled hydrogels provide a conducive matrix for robust biofilm formation by both Gram$^-$ and Gram$^+$ pathogens.

### 3. Validation and Characterization of recolonisation model

*Morphology characterization of recolonized biofilms*

To assess the capacity of the PAs/(A)WF system to support 3D recolonized biofilm formation, a *P. aeruginosa* model was established by connecting a fragment of mature PA/WF biofilm hydrogel to a layer of clean serum-based PA/AWF hydrogel at the bottom of an insert, taking advantage of the bacterium's high motility and robust biofilm-forming ability (Figure. 3a). Alginate hydrogel was used as a control, given its widespread application as a biocompatible scaffold with passive statical encapsulation properties[53,54]. Recolonisation on alginate hydrogel resulted in a sparse, two-dimensional bacterial layer restricted to the material surface, as revealed by SEM (Figure. 3b). In contrast, bacteria recolonizing the PA/AWF hydrogel formed a thick, 3D colonization zone, where bacterial cells were found to intertwine with the fibrous PA/WF networks, suggesting an active structural integration that may facilitate spatial organization during biofilm reformation (Figure. 3b). Notably, recolonized biofilms formed in the GF-AGD/WF group contained more continuous and structured nanofibres than those in the AGD/WF group (Figure. 3b). This enhancement is attributed to the MMP-9-inhibitory function of the GF motif, which may have facilitated improved retention and accumulation of extracellular polymeric substance (EPS) within the matrix. To examine whether 3D recolonized biofilms confer increased antibiotic tolerance, the GF-AGD/WF recolonisation system was exposed to antibiotics for 2 h, after which colony-forming unit (CFU) quantification of the primary biofilm revealed significantly higher CFU counts (Extended Data Fig. 4a). This effect is due to the recolonized biofilm impeding antibiotic penetration and thereby reducing direct activity against the overlying primary biofilm. In conclusion, these observations demonstrate that the GF-AGD/AWF provides a bioactive and structurally supportive microenvironment that promotes 3D recolonized biofilm formation, which subsequently impart unique antibiotic tolerance to the overall biofilm model.

*Transcriptomic analysis of primary and recolonized 3D biofilms*

To assess the global transcriptional differences between primary and recolonized *P. aeruginosa* biofilms, we first performed principal component analysis (PCA). The results revealed a clear separation along PC1 and PC2, which explained 76.73% and 27.59% of the total variance, respectively (Figure 3c), indicating distinct transcriptional landscapes between the two groups. Consistently, Pearson correlation analysis showed high intra-group correlation ($R^2 > 0.95$), while inter-group correlation was markedly lower, confirming the reliability of the transcriptomic data (Figure. 3d). A total of 1,138 differentially expressed genes (DEGs) were identified between primary and recolonized biofilms, including 79 upregulated and 1,059 downregulated transcripts ($|log_2FC| > 1.5$, $FDR < 0.05$). DEG patterns were visualized using hierarchical clustering (Figure. 3e). Gene Ontology (GO) enrichment analysis revealed that DEGs were significantly enriched in biological processes such as cellular processes, translation, and organonitrogen compound biosynthesis (Figure. 3f). Kyoto Encyclopedia of Genes and Genomes (KEGG) analysis further highlighted enriched pathways including ribosome biogenesis, bacterial chemotaxis, RNA degradation, and flagellar assembly (Figure. 3g).

To explore functional shifts in resistance and pathogenicity, we annotated the transcriptomic dataset using the Comprehensive Antibiotic Resistance Database (CARD) and the Virulence Factors Database (VFDB). CARD annotation identified 89 antibiotic resistance-related genes, among which members of the multidrug efflux RND family *muxA*, *mexB*, *mexA*, and *mexX* were significantly downregulated in recolonized biofilms (Figure. 3h), suggesting a reduced efflux-mediated resistance phenotype at this stage. VFDB annotation revealed that several virulence-associated genes were also markedly downregulated in recolonized biofilms, particularly those involved in adhesion, secretion systems, nutrient acquisition, and outer membrane integrity (Figure. 3i). Notably, downregulated transcripts included *OprD* family porins, fibronectin-binding proteins, the type III secretion system (T3SS) master regulator *exsA*, and the metabolic virulence factor *glnA*. This expression pattern suggests that recolonized biofilms may occupy a transitional, low-virulence state. The suppression of outer membrane porins and host matrix-binding proteins likely reflects a transient or unstable adhesion phase following migration to a new surface. Concurrent downregulation of *exsA* implies a deactivation of acute virulence programs, consistent with a shift toward persistence over invasion. Furthermore, the repression of metabolic virulence factors such as *glnA* suggests a metabolically quiescent state, potentially representing an energy-conserving strategy during early reattachment.

In conclusion, these transcriptional alterations delineate a distinct physiological state in recolonized biofilms, characterized by attenuated adhesion, suppressed virulence, and reduced metabolic activity. This likely represents a strategic phenotypic remodelling that facilitates adaptation to new abiotic surfaces during biofilm dispersal and re-establishment. These findings provide mechanistic insight into the dynamic regulation of virulence and resistance during biofilm reattachment and may have important implications for the persistence and recurrence of biofilm infections.

**4. Antibiotic screening kit of polymicrobial biofilm *in vitro***

*PA-SA/GF-AGD/AWF polymicrobial biofilm culture*
Having established the recolonization dynamics, we next investigated the system's capacity to form composite biofilms, a feature critical to polymicrobial chronic wound infections. *P. aeruginosa* and *S. aureus* were selected as representative model organisms, corresponding to common Gram$^-$ rods and Gram$^+$ cocci frequently isolated from chronic wound infections[55]. The PA/AWF hydrogel supported bacterial growth and recolonisation, enabling the expansion of biofilms over 48 hours and leading to the formation of macroscale, centimetre-scale biofilm structures. By varying the initial inoculation ratios (*P. aeruginosa*: *S. aureus* = 1:100, 1:1, and 100:1), we generated *S. aureus*-dominant (S-D), balanced (B), and *P. aeruginosa*-dominant (P-D) biofilms respectively (Figure. 4a, Extended Data Fig. 4b). These results demonstrate that the GF-AGD/AWF hydrogel platform can reliably sustain the co-culture of Gram$^-$ and Gram$^+$ species, enabling tuneable biofilm composition and structural development that closely mimic polymicrobial communities in chronic wounds.

*Antibiotic screening kit vs. clinical method*

We next evaluated the antimicrobial responsiveness of the B-biofilm model using three clinically relevant antibiotics: polymyxin B (PB, active against Gram$^-$ bacteria), vancomycin (VAN, active against Gram$^+$ bacteria), and ciprofloxacin (Cip, a broad-spectrum antibiotic). Mature biofilms were treated with antibiotic/AWF solutions for 24 hours, after which the OD$_{600}$ of the surrounding medium was measured to assess bacterial growth and regrowth (Figure. 4b). For B-biofilm, compared to untreated control group, Cip-treated group and PB-treated group showed inhibition of both growth and regrowth to the dual-species biofilm at relatively low (8 μg/mL) and moderate (128 μg/mL) concentrations, respectively. In contrast, VAN-treated group suppressed initial bacterial growth at 64 μg/mL, but it failed to prevent regrowth with the minimum biofilm eradication concentration (MBEC) exceeding 1024 μg/mL. Stereomicroscopic imaging of the post-treatment B-biofilms revealed distinct morphological responses. Cip completely disrupted the biofilm macrostructure, while PB only disrupted central biofilm structure but preserved the integrity of the marginal biofilm. VAN, however, had minimal impact on the biofilm's structural integrity (Figure. 4c). We believe that the differential efficacies of the two narrow-spectrum antibiotics against dual-species biofilms arise from the distinct structural roles of their target organisms. In B-biofilms, *P. aeruginosa* functions as the principal structural maintainer, explaining why PB displayed stronger biofilm-disrupting activity than VAN. Consistently, VAN effectively disrupted S-D biofilms but was ineffective against P-D biofilms, whereas PB produced the opposite pattern (Extended Data Fig. 4c). Live/dead assay and subsequent quantitative analysis confirmed the superior bactericidal activity of Cip, followed by PB, and VAN exhibited the lowest efficacy (Figure. 4c and 4e). SEM revealed antibiotics-specific ultrastructural responses of *P. aeruginosa* and *S. aureus* in B-biofilm to antibiotic treatment. Following exposure to PB, *P. aeruginosa* cells exhibited pronounced outer-membrane protrusions[56], indicative of membrane stress and potential lipid rearrangement (Figure. 4d). In contrast, *S. aureus* treated with VAN displayed extensive membrane perforation and surface collapse[57], consistent with the disruption of peptidoglycan integrity (Figure. 4d). These results demonstrate that within our GF-AGD/AWF dual-species biofilm model, Cip displays the most potent bactericidal activity, followed by PB, while VAN is markedly less effective in eradicating and disrupting mature biofilms *in vitro*.

To validate the predictive reliability of our *in vitro* antibiotic screening model, we conducted parallel antimicrobial susceptibility testing using standard agar diffusion assays as clinical technology control group. The inhibition zone diameters produced by the three antibiotics on CAMHA inoculated with *S. aureus* and *P. aeruginosa* validated the reliability of our GF–AGD/AWF *in vitro* antibiotic screening model, as they exhibited the same ranking pattern (Figure. 4f). Taken together, our results validated the reliability of the GF-AGD/WF antibiotic screening model, which yields comparable outcomes to the conventional disc diffusion method. Moreover, by eliminating the requirement for pathogen isolation and culture, this strategy provides a simplified and potentially faster alternative for clinical antibiotic susceptibility assessment.

**5. The efficacy of antibiotic screening kit demonstrated by *in vivo* chronic wound infection model**

*Establishment of a polymicrobial biofilm chronic wound infection in vivo model*

To validate our GF-AGD/WF *in vitro* model, we compared its biofilm morphology and antibiotic response to an *in vivo* benchmark. We established this benchmark by generating a balanced, co-inoculated *P. aeruginosa* and *S. aureus* (1:1) biofilm in a rat full-thickness wound model, which developed through sequential colonisation, dispersal, and maturation phases, recapitulating the biological progression observed in GF-AGD/AWF model *in vitro* (Figure. 5a). SEM further revealed that the *in vivo* biofilms exhibited a microstructure closely resembling that of the GF-AGD/WF biofilms, with bacterial aggregates embedded within a network of nanofibres (Figure. 5a). These *in vivo* biofilms were then treated with the same antibiotics used in our prior *in vitro* studies. Cip markedly accelerated wound closure compared with PB-treated, VAN-treated and control group, indicating superior efficacy against the dual-species biofilm, as assessed after 48 hours of biofilm maturation and subsequent antibiotic treatment (Figure. 5b). Haematological analysis demonstrated an absence of overt drug-related toxicity, as measured by standard blood tests (Figure. 5c). Body weight fluctuations were maintained within 10% throughout the experimental period for all animals, confirming overall health and compliance with international animal welfare standards (Figure. 5d). Notably, Cip treatment resulted in the lowest leukocyte counts and plasma IL-1β and IL-6 concentrations, reflecting reduced systemic inflammation following effective biofilm clearance, as quantified by haematological and cytokine assays (Figures. 5c, 5e, and 5f). Haematoxylin and Eosin (H&E) staining revealed that all infected groups exhibited thick scab layers, indicative of microbial colonization. Among the treatment groups, Cip-treated group showed the lowest level of inflammatory cell infiltration, whereas wounds of PB-treated group and VAN-treated group exhibited more pronounced infiltration (Figure 6a). These results validate the results of GF-AGD/AWF *in vitro* antibiotic screening kit evaluation of antibiotic efficacy.

*Similarity of microbial biogeography in co-assembled in vitro and in vivo models*

To determine the structural and functional parallels between the *in vivo* and *in vitro* biofilms, we analysed their composition and spatial organization using Gram staining, scanning electron microscopy (SEM), and fluorescence *in situ* hybridization (FISH). Gram staining confirmed that bacteria were predominantly localized within the scab, forming dense aggregates in the untreated control group (within the yellow dashed line). Wounds of VAN-treated group displayed bacterial aggregates of approximately 10 μm in diameter (red arrows), while wounds of PB-treated group showed smaller aggregates (~5 μm, green arrows). In contrast, no bacterial aggregates were observed in the Cip-treated group (Figure. 6b). An absence of biofilm aggregates in both the Cip-treated and blank groups was revealed by SEM (Figure. 6c). The extracellular matrix in these groups exhibited a loose fibrous meshwork, potentially representing the fibronectin-based wound bed architecture synthesized by the host. Meanwhile, dense biofilm structures with shortened fibrous elements were observed in the PB-treated, VAN-treated, and control groups (Figure. 6c). This morphological alteration resulted from interactions between biofilm growth and host-derived proteins that reshape the microstructure of the wound bed. Notably, the dominant bacterial species within the biofilm varied among treatment groups: cocci (*S. aureus*) predominated in the PB-treated group, whereas bacilli (*P. aeruginosa*) dominated in the VAN-treated group, aligning with the known antimicrobial spectra of the respective antibiotics. These *in vivo* findings were consistent with the *in vitro* results, demonstrating a relative efficacy ranking of Cip > PB > VAN. FISH further confirmed microbial distribution and species-specific biofilm responses. In untreated biofilms, strong co-localization of *P. aeruginosa* and *S. aureus* was observed (Figure. 6d). Cip effectively eradicated both bacterial species and disrupted biofilm structure. PB resulted in fewer and smaller aggregates (<5 μm$^2$, Figure. 6d and 6e), whereas VAN selectively cleared *S. aureus* but allowed *P. aeruginosa* to persist, forming larger aggregates (5 μm$^2$~10μm$^2$, Figure. 6d and 6f). Taken together, these findings suggest a structural role for *P. aeruginosa* in maintaining biofilm integrity; the absence of *P. aeruginosa* may destabilise the biofilm matrix and facilitate the clearance of co-residing *S. aureus*. Collectively, these *in vivo* findings validate the predictive power of our GF-AGD/AWF co-assembling *in vitro* model. The antibiotic efficacy rankings established through *in vitro* screening were accurately recapitulated in the *in vivo* wound infection model.

**Discussion**

In this study, we developed a PA–based co-assembly system with WF to recapitulate the essential features of *in vivo* infectious biofilms, including their complex nutritional component, 3D architecture, and dispersal–recolonisation dynamics. Notably, the PA/AWF *in vitro* biofilm model exhibited a high degree of concordance with *in vivo* composite biofilms in both antibiotic efficacy profiles and microbial spatial responses. This strong consistency underscores the translational potential of our model for guiding the preclinical evaluation and optimisation of antibiotic strategies against complex and dynamic biofilm-associated infections.The main innovations and strengths of this work include: (1) the design of an original PA-GF sequence with MMP-9 inhibitory activity; (2) the successful *in vitro* validation of the biofilm recolonisation process; (3) the construction of a tunable *in vitro* composite biofilm model that enables controlled investigation of multispecies infections; (4) the reproduction of antibiotic response patterns consistent with *in vivo* observations, supporting its predictive relevance for drug screening. Collectively, this work establishes a biomimetic framework for reconstructing and analysing infection-relevant biofilms, offering a versatile experimental platform that bridges *in vitro* biofilm modelling and clinical decision-making in precision antibiotic therapy.

**Methods**

**Synthesis and characterization of PAs**

The AGD and GF PAs were obtained from GL Biochem (China), and their molecular sequences are illustrated in Figure 1a. These PAs were synthesised using solid-phase peptide synthesis and subsequently characterised by reverse-phase high-performance liquid chromatography (RP-HPLC) and electrospray ionisation mass spectrometry (ESI-MS).

**Microbial strains and culture conditions**

The *P. aeruginosa* ATCC 27853 and *S. aureus* SH1000 strains were routinely cultured in lysogeny broth (LB; Hopebio, Qingdao, China) for 16 h at 37 °C with shaking at 250 rpm. For biofilm formation, 16 h planktonic cultures of both strains were harvested by centrifugation (10,000 × g), washed twice with phosphate-buffered saline (1× PBS), and resuspended in artificial wound fluid (AWF) at the desired cell densities (colony-forming units per millilitre, CFU mL$^{-1}$). For selective microbial isolation, *Pseudomonas* isolation agar with nalidixic acid (CNA; Hopebio) and Baird-Parker agar (BPA; Hopebio) were used for *P. aeruginosa* and *S. aureus*, respectively.

**Dynamic light scattering (DLS)**

The particle size distributions of individual PAs and their co-assemblies were determined by Nanolink instrument (Linkoptik, China). Prior to measurement, the samples were diluted to 0.05 wt% with distilled water, after which the PAs were mixed with WF or AWF at a 1:1 volume ratio to induce co-assembly.

**Electrophoretic light scattering (ELS)**

Zeta potentials of PA (0.01%), AWF (0.05%), and WF (0.05%) solutions were determined at 25 °C using a NanoLink zeta potential analyser (linkoptik). All samples were prepared in ultrapure water and allowed to equilibrate for 10 min at the measurement temperature prior to analysis.

**Circular dichroism (CD)**

CD spectra of PA, AWF, and WF solutions (0.05 wt%) were acquired using a spectrometer (JASCO, Japan). Each sample was prepared in a 0.01 mm pathlength cuvette, and ten consecutive scans were recorded between 190 and 260 nm at a rate of 50 nm min$^{-1}$. The resulting spectra were subsequently processed with a simple moving-average algorithm to minimise noise.

**Transmission electron microscopy (TEM)**

For TEM analysis, 200-mesh carbon-coated copper grids were loaded with PA–AWF and PA–WF co-assemblies suspensions and allowed to stand for 5 min. The samples were subsequently stained with 2 wt% uranyl acetate for 30 s, gently rinsed with ultrapure water for 30 s, and then left to air-dry at ambient temperature for 24 h. Bright-field TEM images were captured using an HT7800 instrument (HITACHI, Japan) operated at an accelerating voltage of 80 kV.

**Scanning electron microscopy (SEM)**

For morphological observation, the PA–AWF and PA–WF co-assemblies were first fixed in 4% paraformaldehyde (PFA; Servicebio, China) for 10 min, followed by graded ethanol dehydration (70%, 80%, 90%, 96%, and 100%) at room temperature. After critical-point drying (HARVENT, China), the dried specimens were sputter-coated with gold and visualised using a GeminiSEM 300 scanning electron microscope (ZEISS, Germany).

**Small-angle neutron scattering (SANS)**

An equal volume of $D_2O$ was added to the AWF and WF solutions to obtain 50% $D_2O$/50% $H_2O$ mixtures. PAs were dissolved in $D_2O$ at a concentration of 2 wt%, and mixed solutions of AWF/WF (50% $D_2O$/50% $H_2O$) and PAs (2 wt%) were prepared for small-angle neutron scattering (SANS) measurements. Experiments were conducted on the BL01 SANS instrument at the China Spallation Neutron Source (China). Individual component solutions (1 mL) were loaded into 1 mm path-length Hellma quartz cuvettes, whereas mixed samples were placed in demountable cuvettes of the same thickness. All cuvettes were secured in aluminium holders inside a sealed, temperature-controlled chamber maintained at 25 °C. Each measurement was completed within approximately 30 min. All scattering profiles were normalized to sample transmission, background-subtracted using a $D_2O$-filled quartz cell, and corrected for detector linearity and efficiency through the instrument-specific data processing software. Neutron beams with wavelengths between 1.1 and 9.8 Å were defined using a double-disc bandwidth chopper and collimated to the sample via a dual-aperture system. The sample-to-detector distance was set to 5 m with an 8 mm sample aperture. A two-dimensional $^3$He tube array detector enabled data collection over a Q range of 0.004–0.7 Å$^{-1}$. For each sample, including the empty holder and cell, scattering signals were recorded for approximately 60 min. The resulting data were converted to absolute units following standard transmission correction and calibration procedures.

**Molecular docking**

The crystal structure of MMP-9 was retrieved from the UniProt database (PDB ID: 1GKC) and subsequently repaired to restore any missing atoms. The peptide structure was modelled using AlphaFold3 and further refined in Discovery Studio Visualizer 2024. Both the protein receptor and peptide ligand were processed individually using AutoDockTools 4, and the prepared files were saved in pdbqt format. Molecular docking was performed using AutoDock Vina 1.2.5, with the processed protein serving as the receptor and the peptide as the ligand. The grid spacing and exhaustiveness parameters were maintained at their default values of 0.375 and 8, respectively. The docking procedure was executed by importing the corresponding pdbqt files of the receptor and ligand into AutoDock Vina under the specified parameters. Following docking, all resulting conformations were ranked according to their binding affinity scores, and the top-ranked complex was selected for subsequent structural analysis.

**MMP-9 inhibition assay**

Recombinant MMP-9 (ACRO Biosystems, China) was activated with 1 mM p-aminophenylmercuric acetate (APMA, Sigma, USA) in TCNB buffer (Sigma) at concentrations of 50 μg/mL and 100 μg/mL, respectively, at 37 °C. The inhibitory effect of the GF sequence on MMP-9 activity was evaluated using the EnzChek® Gelatinase/Collagenase Assay Kit (Thermo Fisher Scientific, USA) following the manufacturer's instructions.

**Extraction of WF from infected wounds**

WF were collected from clinical dressings under sterile conditions. The outer dressing layer was removed to expose the inner dressing. Using sterile forceps, the inner layer soaked with WF was transferred into sterile 50 mL centrifuge tubes and sealed. Within a biosafety cabinet, the collected dressings were cut into small pieces with sterile scissors and incubated with an equal volume of sterile saline (1:1 v/v) at 4 °C for 24 h. After incubation, the initially clear saline became visibly turbid, indicating the release of microbial components and WF soluble factors. The incubation solution was recovered using sterile Pasteur pipettes and filtered through a 70 μm cell strainer to remove residual fabric debris and impurities, yielding the infected wound exudate containing pathogenic microorganisms. A portion of the exudate was subsequently sterilized by dual filtration through 0.22 μm syringe filters to obtain sterile WF.

**AWF preparation**

AWF was prepared to replicate the biochemical composition and ionic environment of chronic wound exudate. Sodium chloride (0.36%), glucose (0.1%), tyrosine (0.03%), sodium carbonate (0.05%), sodium citrate (0.02%), ferrous sulphate heptahydrate (0.02%), magnesium chloride hexahydrate (0.02%), sodium lactate (0.1%), calcium chloride dihydrate (0.01%), potassium dihydrogen phosphate (0.054%), and urea (0.01%) were sequentially dissolved in ultrapure water under gentle stirring. The solution was then supplemented with thiamine hydrochloride (0.02%) and nicotinic acid (0.02%) as essential cofactors. After complete dissolution, 20% (v/v) human serum was added under aseptic conditions. The pH of the mixture was adjusted to 7.4 ± 0.1 using 0.1 M NaOH or HCl. The resulting solution was sterilized by filtration through a 0.22 μm membrane, aliquoted, and stored at 4 °C for short-term use (≤7 days).

### Sodium dodecyl sulphate-polyacrylamide gel electrophoresis (SDS-PAGE)

After gelation for 1 h, PA–WF hydrogels were washed with PBS for 10 min to remove unbound WF components. The hydrogels were then disrupted by pipetting and vortexing, followed by a 10-fold dilution in PBS. Aliquots (15 μL) were mixed with 5 μL of loading buffer (Sigma), heated at 90 °C for 5 min, and loaded onto pre-cast NuPAGE Bis-Tris gels (ACE Biotechnology, China). A pre-stained protein ladder (10–180 kDa; Thermo Fisher Scientific) was used as a molecular weight reference. Electrophoresis was performed for 1 h at 120 V in NuPAGE MES SDS running buffer, followed by Coomassie Brilliant Blue staining (Beyotime, China) and imaging using a gel documentation system (BLT-IMAGING, China).

### Library Construction and Sequencing

Total RNA was extracted from the samples using the cetyltrimethylammonium bromide (CTAB) method, and residual genomic DNA was removed. Only high-quality RNA was used for library preparation. Ribosomal RNA (rRNA) was depleted using the RiboCop rRNA Depletion Kit for Mixed Bacterial Samples (Lexogen, USA) instead of poly(A) selection to ensure unbiased transcript coverage. Following rRNA removal, the remaining RNA was fragmented into approximately 300-nucleotide fragments using a fragmentation buffer. First-strand cDNA synthesis was performed using random hexamer primers, followed by second-strand synthesis during which dUTP was incorporated in place of dTTP to maintain strand specificity. The resulting double-stranded cDNA underwent end-repair, phosphorylation, and adenylation according to Illumina's standard library preparation protocol. Strand-specific RNA-seq libraries were generated using the Illumina Stranded mRNA Prep, Ligation Kit (Illumina, San Diego, CA, USA), and paired-end sequencing was performed on the Illumina NovaSeq X Plus platform. Image analysis, base calling, and quality scoring were conducted using the Illumina pipeline. Low-quality reads, adaptor-contaminated reads, and sequences containing more than 10% ambiguous bases (N) were removed to obtain high-quality clean reads for downstream analyses.

### Bioinformatics Analysis

Sequencing data generated from the Illumina platform were processed and analysed using the Majorbio Cloud Platform (https://cloud.majorbio.com/) provided by Shanghai Majorbio Bio-Pharm Technology Co., Ltd. (Shanghai, China). High-quality clean reads from each sample were aligned to the reference genome using Bowtie2. To assess potential rRNA contamination, 10,000 raw reads were randomly selected from each sample and aligned to the Rfam database (http://rfam.xfam.org/) using the DIAMOND algorithm. Based on the annotation results, the proportion of reads mapped to rRNA sequences was calculated to estimate the rRNA contamination level in each sample.

**Expression analysis**

Gene and transcript abundances from both single-end and paired-end RNA-Seq datasets were quantified using RSEM. For each dataset and alignment protocol, differentially expressed genes (DEGs) were identified using the DESeq2 package, based on normalized read counts and statistical significance thresholds.

**DEGs Gene Ontology (GO) and Kyoto Encyclopaedia of Genes and Genomes (KEGG) enrichment analysis**

GO enrichment analysis was performed using the GOATOOLS package with Fisher's exact test. To control the false positive rate, four multiple testing correction methods including Bonferroni, Holm, Sidak, and false discovery rate (FDR) were applied to adjust the P-values. GO terms with adjusted P-values (< 0.05) were considered significantly enriched. KEGG pathway enrichment analysis was conducted using R scripts, following the same statistical principle as the GO analysis. Enrichment significance was determined by Fisher's exact test, and the same four correction methods were used to control the false discovery rate. KEGG pathways with adjusted P-values (< 0.05) were defined as significantly enriched among the differentially expressed genes.

**DEGs Comprehensive Antibiotic Resistance Database (CARD) and Virulent Factor Database (VFDB) annotation analysis**

DEGs were functionally annotated against the CARD and the VFDB. The protein sequences of DEGs were aligned to both databases using the DIAMOND program with a sequence identity threshold of ≥ 40%, alignment coverage of ≥ 70%, and an E-value ≤ 1e−5. Based on the alignment results, antibiotic resistance-related genes and their resistance mechanisms, including antibiotic efflux, target modification, and enzyme-mediated inactivation, were identified and classified according to resistance types. Similarly, VFDB annotation was performed to identify potential virulence-associated genes such as adhesins, secretion systems, toxins, and immune evasion factors. The CARD and VFDB annotation results were integrated with the expression profiles of DEGs to further analyse the variation in resistance and virulence-related gene expression across primary and recolonisation biofilms.

**Antibiotics susceptibility test**

All antibiotics were purchased from MCE (MedChemExpress, China). The minimum biofilm inhibitory concentration (MBIC), minimum biofilm eradication concentration (MBEC), and agar diffusion assays were performed following previously reported protocols[58].

**colony forming unit (CFU) quantification**

To quantify viable bacteria in different treatment biofilm groups, CFU counting was performed using a spot plating method. Briefly, each sample was thoroughly mixed and serially diluted tenfold in sterile PBS. Aliquots of 5 μL from each dilution were spotted onto selective agar plates, including *Pseudomonas* CN agar for *P. aeruginosa* and Baird-Parker agar for *S. aureus*. The spots were allowed to absorb and dry on the agar surface, and the plates were incubated at 37 °C for 16–18 h. Visible colonies within each spot were counted, and CFU per millilitre or per gram of sample was calculated according to the dilution factor.

**Live and dead staining**

Biofilm viability was evaluated using a LIVE/DEAD™ BacLight™ Bacterial Viability Kits (Thermo Fisher Scientific) following the manufacturer's protocol. Briefly, biofilms were cultured on PAs-based hydrogels or in control wells for the indicated durations. The supernatant was removed, and the samples were gently rinsed twice with sterile PBS to eliminate non-adherent cells. A staining solution containing SYTO-9 and propidium iodide (PI) at final concentrations of 6 μM and 30 μM, respectively, was prepared in PBS and added to fully cover the biofilm surface. Samples were incubated in the dark at room temperature for 15 min, followed by gentle washing with PBS to remove excess dye. For hydrogel-embedded biofilms, confocal z-stack images were acquired using a laser scanning confocal microscope (Leica TCS SP8, Germany) equipped with 488 nm and 561 nm excitation lasers for SYTO-9 (live cells, green) and PI (dead cells, red) channels, respectively. Images were processed using LAS X and ImageJ software to reconstruct 3D biofilm structures and to quantify live/dead cell ratios based on fluorescence intensity.

**Animal experiments**

Sprague-Dawley rats (male, 8 weeks old; Wuhan Boerfu Biotechnology Co., Ltd., China) were randomly divided into five groups (n = 3 per group): untreated (U-T), ciprofloxacin-treated (C-T), polymyxin B-treated (P-T), vancomycin-treated (V-T), and blank control. The dorsal hair of each rat was shaved, and the skin was disinfected with 75% ethanol. Following anaesthesia induction with inhaled isoflurane, full-thickness excisional wounds (8 mm in diameter) were created on the dorsal skin using a sterile biopsy punch. Each wound was inoculated with 100 μL of a bacterial suspension containing *P. aeruginosa* ($1 \times 10^7$ CFU) and *S. aureus* ($1 \times 10^7$ CFU), then covered with a sterile dressing. Three days after inoculation, composite biofilm infections were established on the wound surface. Antibiotics were administered intraperitoneally once daily for three consecutive days. Wound healing progression was monitored and recorded daily by digital photography.

**Histological staining**

Rats were sacrificed on day 3 for histological and Gram staining analyses of wound tissues. Skin samples were fixed in 10% neutral buffered formalin, dehydrated through graded ethanol series, embedded in paraffin, and sectioned at a thickness of 3–4 μm. After deparaffinization and rehydration, the sections were stained with haematoxylin and eosin (H&E) for histological examination and with Gram stain for biofilm visualization. The stained slides were observed under a light microscope.

**Plasma cytokine and haematological analysis**

On day 3 post-infection, rats were sacrificed and whole blood was collected via retro-orbital puncture. Plasma was separated by centrifugation at 3000 × g for 10 min at 4 °C and stored at −80 °C. Levels of interleukin-1β (IL-1β) and interleukin-6 (IL-6) in plasma were quantified using ELISA kits (Wuhan Boerfu Biotechnology Co., Ltd., China) according to the manufacturer's instructions. Absorbance was measured at 450 nm using a microplate reader (BioTek, USA). Haematological parameters, including total white blood cell, red blood cell, haemoglobin, and platelet counts, were determined using an automated haematology analyser (Sysmex, Japan).

**Fluorescence *in situ* hybridization (FISH)**

Fluorescence *in situ* hybridization (FISH) was performed as previously described[59]. The *Pseudomonas aeruginosa* species-specific probe (Boerfu, China) had the sequence GGACGTTATCCCCCACTAT[60], with a hybridization temperature of 50 °C. The *Staphylococcus aureus* species-specific probe (Boerfu, China) had the sequence AGAGAAGCAAGCTTCTCGTCCG[61], with a hybridization temperature of 46 °C. Blank control and U-T group ware included as positive and negative control. Tissue specimens for FISH analysis were selected by an experienced histopathologist, focusing on regions showing histological signs of inflammation. Bacterial colonies located within the focal plane of tissue structures were identified as positive only when the corresponding control samples yielded expected results.

**Statistical analysis**

Statistical analyses were performed using one-way or two-way ANOVA followed by Tukey's post-hoc test in GraphPad Prism (version 10, GraphPad Software, USA). Differences were considered statistically significant at $P < 0.05$. Significance levels were denoted as follows: ns, $P > 0.05$; $P < 0.05$; *$P < 0.01$; **$P < 0.001$; and ***$P < 0.0001$.

**References**

1. Falanga, V. *et al.* Chronic wounds. *Nat Rev Dis Primers* **8**, 50 (2022).

2. Uberoi, A., McCready-Vangi, A. & Grice, E. A. The wound microbiota: microbial mechanisms of impaired wound healing and infection. *Nat Rev Microbiol* **22**, 507–521 (2024).
3. Chung, J. Y. *et al.* Microbubble-Controlled Delivery of Biofilm-Targeting Nanoparticles to Treat MRSA Infection. *Adv Funct Materials* **35**, 2508291 (2025)
4. Hall-Stoodley, L., Costerton, J. W. & Stoodley, P. Bacterial biofilms: from the Natural environment to infectious diseases. *Nat Rev Microbiol* **2**, 95–108 (2004).
5. Flemming, H.-C. *et al.* Microbial extracellular polymeric substances in the environment, technology and medicine. *Nat Rev Microbiol* **23**, 87–105 (2025).
6. Flemming, H.-C. *et al.* The biofilm matrix: multitasking in a shared space. *Nat Rev Microbiol* **21**, 70–86 (2023).
7. Louis, M. *et al. Pseudomonas aeruginosa* Biofilm Dispersion by the Human Atrial Natriuretic Peptide. *Advanced Science* **9**, 2103262 (2022).
8. Kearney, K. J., Ariëns, R. A. S. & Macrae, F. L. The Role of Fibrin(ogen) in Wound Healing and Infection Control. *Semin Thromb Hemost* **48**, 174–187 (2022).
9. Yi, J. *et al.* Orthopedic Implant Infection Management: Prevention, Barrier Breakthrough, and Immunomodulation. *ACS Nano* **19**, 27009–27032 (2025).
10. Rumbaugh, K. P. & Whiteley, M. Towards improved biofilm models. *Nat Rev Microbiol* **23**, 57–66 (2025).
11. Wu, B. *et al.* Human organoid biofilm model for assessing antibiofilm activity of novel agents. *npj Biofilms Microbiomes* **7**, 8 (2021).
12. Koo, H., Allan, R. N., Howlin, R. P., Stoodley, P. & Hall-Stoodley, L. Targeting microbial biofilms: current and prospective therapeutic strategies. *Nat Rev Microbiol* **15**, 740–755 (2017).
13. Lebeaux, D., Ghigo, J.-M. & Beloin, C. Biofilm-Related Infections: Bridging the Gap between Clinical Management and Fundamental Aspects of Recalcitrance toward Antibiotics. *Microbiol Mol Biol Rev* **78**, 510–543 (2014).
14. Wu, Y. *et al.* Cooperative microbial interactions drive spatial segregation in porous environments. *Nat Commun* **14**, 4226 (2023).
15. Cometta, S., Hutmacher, D. W. & Chai, L. *In vitro* models for studying implant-associated biofilms - A review from the perspective of bioengineering 3D microenvironments. *Biomaterials* **309**, 122578 (2024).
16. Ning, E. *et al.* 3D bioprinting of mature bacterial biofilms for antimicrobial resistance drug testing. *Biofabrication* **11**, 045018 (2019).
17. Sun, Y., Dowd, S. E., Smith, E., Rhoads, D. D. & Wolcott, R. D. *In vitro* multispecies Lubbock chronic wound biofilm model. *Wound Repair Regeneration* **16**, 805–813 (2010).
18. Di Fermo, P. *et al.* Antimicrobial Peptide L18R Displays a Modulating Action against Inter-Kingdom Biofilms in the Lubbock Chronic Wound Biofilm Model. *Microorganisms* **9**, 1779 (2021).
19. Sauer, K. *et al.* The biofilm life cycle: expanding the conceptual model of biofilm

formation. *Nat Rev Microbiol* **20**, 608–620 (2022).

20. Trego, A. C. *et al.* Size Shapes the Active Microbiome of Methanogenic Granules, Corroborating a Biofilm Life Cycle. *mSystems* **5**, e00323-20 (2020).

21. Kaplan, J. B. Biofilm Dispersal: Mechanisms, Clinical Implications, and Potential Therapeutic Uses. *J Dent Res* **89**, 205–218 (2010).

22. Chua, S. L. *et al.* Dispersed cells represent a distinct stage in the transition from bacterial biofilm to planktonic lifestyles. *Nat Commun* **5**, 4462 (2014).

23. Rumbaugh, K. P. & Sauer, K. Biofilm dispersion. *Nat Rev Microbiol* **18**, 571–586 (2020).

24. Weller, J. *et al.* The role of bacterial corrosion on recolonization of titanium implant surfaces: An *in vitro* study. *Clin Implant Dent Rel Res* **24**, 664–675 (2022).

25. Álvarez, S., Leiva-Sabadini, C., Schuh, C. M. A. P. & Aguayo, S. Bacterial adhesion to collagens: implications for biofilm formation and disease progression in the oral cavity. *Critical Reviews in Microbiology* **48**, 83–95 (2021).

26. Bhattacharya, M. & Horswill, A. R. The role of human extracellular matrix proteins in defining *Staphylococcus aureus* biofilm infections. *FEMS Microbiol Rev* **48**, fuae002 (2024).

27. McDougald, D., Rice, S. A., Barraud, N., Steinberg, P. D. & Kjelleberg, S. Should we stay or should we go: mechanisms and ecological consequences for biofilm dispersal. *Nat Rev Microbiol* **10**, 39–50 (2011).

28. Hwang, G. *et al.* Catalytic antimicrobial robots for biofilm eradication. *Sci. Robot.* **4**, eaaw2388 (2019).

29. Ma, Y., Deng, Y., Hua, H., Khoo, B. L. & Chua, S. L. Distinct bacterial population dynamics and disease dissemination after biofilm dispersal and disassembly. *The ISME Journal* **17**, 1290–1302 (2023).

30. Cometta, S., Hutmacher, D. W. & Chai, L. *In vitro* models for studying implant-associated biofilms - A review from the perspective of bioengineering 3D microenvironments. *Biomaterials* **309**, 122578 (2024).

31. Postek, W., Staśkiewicz, K., Lilja, E. & Wacław, B. Substrate geometry affects population dynamics in a bacterial biofilm. *Proc. Natl. Acad. Sci. U.S.A.* **121**, e2315361121 (2024).

32. Cont, A., Vermeil, J. & Persat, A. Material Substrate Physical Properties Control *Pseudomonas aeruginosa* Biofilm Architecture. *mBio* **14**, e03518-22 (2023).

33. Datta, D. *et al.* Phenotypically complex living materials containing engineered cyanobacteria. *Nat Commun* **14**, 4742 (2023).

34. Makam, P. & Gazit, E. Minimalistic peptide supramolecular co-assembly: expanding the conformational space for nanotechnology. *Chem. Soc. Rev.* **47**, 3406–3420 (2018).

35. Zhou, J. *et al.* Self-Healable Organogel Nanocomposite with Angle-Independent Structural Colors. *Angew Chem Int Ed* **56**, 10462–10466 (2017).

36. Chakraborty, S., Khamrui, R. & Ghosh, S. Redox responsive activity regulation in exceptionally stable supramolecular assembly and co-assembly of a protein. *Chem. Sci.* **12**,

1101–1108 (2021).

37. Kaizerman-Kane, D. *et al.* pH-Responsive Pillar[6]arene-based Water-Soluble Supramolecular Hexagonal Boxes. *Angew Chem Int Ed* **58**, 5302–5306 (2019).

38. Dutta, R., Sil, S., Kundu, S., Nandi, S. & Sarkar, N. Multi-stimuli responsive fabrication of supramolecular assemblies using ionic self-assembly approach. *Journal of Molecular Liquids* **286**, 110861 (2019).

39. Chen, Y., Szkopek, T. & Cerruti, M. Supramolecular temperature responsive assembly of polydopamine reduced graphene oxide. *Mater. Horiz.* **10**, 2638–2648 (2023).

40. Wu, Y. *et al.* Co-assembling living material as an *in vitro* lung epithelial infection model. *Matter* **7**, 216–236 (2023).

41. Padilla-Lopategui, S. *et al.* Biocooperative Regenerative Materials by Harnessing Blood-Clotting and Peptide Self-Assembly. *Advanced Materials* **36**, 2407156 (2024).

42. Thaarup, I. C. & Bjarnsholt, T. Current *In Vitro* Biofilm-Infected Chronic Wound Models for Developing New Treatment Possibilities. *Advances in Wound Care* **10**, 91–102 (2020).

43. Diban, F. *et al.* Biofilms in Chronic Wound Infections: Innovative Antimicrobial Approaches Using the *In Vitro* Lubbock Chronic Wound Biofilm Model. *IJMS* **24**, 1004 (2023).

44. Maslova, E., Eisaiankhongi, L., Sjöberg, F. & McCarthy, R. R. Burns and biofilms: priority pathogens and *in vivo* models. *npj Biofilms Microbiomes* **7**, 73 (2021).

45. Thaarup, I. C., Iversen, A. K. S., Lichtenberg, M., Bjarnsholt, T. & Jakobsen, T. H. Biofilm Survival Strategies in Chronic Wounds. *Microorganisms* **10**, 775 (2022).

46. Rodrigues, M., Kosaric, N., Bonham, C. A. & Gurtner, G. C. Wound Healing: A Cellular Perspective. *Physiological Reviews* **99**, 665–706 (2019).

47. Uberoi, A., McCready-Vangi, A. & Grice, E. A. The wound microbiota: microbial mechanisms of impaired wound healing and infection. *Nat Rev Microbiol* **22**, 507–521 (2024).

48. Puthia, M. *et al.* A dual-action peptide-containing hydrogel targets wound infection and inflammation. *Sci. Transl. Med.* **12**, eaax6601 (2020).

49. Hernandez-Guillamon, M. *et al.* Sequential Amyloid-β Degradation by the Matrix Metalloproteases MMP-2 and MMP-9. *Journal of Biological Chemistry* **290**, 15078–15091 (2015).

50. Wysocki, A. B., Staiano-Coico, L. & Grinnell, F. Wound Fluid from Chronic Leg Ulcers Contains Elevated Levels of Metalloproteinases MMP-2 and MMP-9. *Journal of Investigative Dermatology* **101**, 64–68 (1993).

51. Rayment, E. A., Upton, Z. & Shooter, G. K. Increased matrix metalloproteinase-9 (MMP-9) activity observed in chronic wound fluid is related to the clinical severity of the ulcer. *Br J Dermatol* **158**, 951–961 (2008).

52. Koivunen, E. *et al.* Tumor targeting with a selective gelatinase inhibitor. *Nat Biotechnol* **17**, 768–774 (1999).

53. Sønderholm, M. *et al. Pseudomonas aeruginosa* Aggregate Formation in an Alginate Bead

Model System Exhibits *In Vivo*-Like Characteristics. *Appl Environ Microbiol* **83**, (2017).

54. Kashi, P. A., Bachlechner, C., Huc-Mathis, D., Jäger, H. & Shahbazi, M. A porous 3D biofilm-inspired alginate/gellan hydrogel fabricated via dual-wavelength UV-crosslinking printer: Structural and rheological properties. *Carbohydrate Polymers* **370**, 124246 (2025).

55. Uberoi, A., McCready-Vangi, A. & Grice, E. A. The wound microbiota: microbial mechanisms of impaired wound healing and infection. *Nat Rev Microbiol* **22**, 507–521 (2024).

56. Borrelli, C. *et al.* Polymyxin B lethality requires energy-dependent outer membrane disruption. *Nat Microbiol* **10**, 2919–2933 (2025).

57. Salamaga, B. *et al.* Demonstration of the role of cell wall homeostasis in *Staphylococcus aureus* growth and the action of bactericidal antibiotics. *Proc. Natl. Acad. Sci. U.S.A.* **118**, e2106022118 (2021).

58. Åhman, J., Matuschek, E. & Kahlmeter, G. Evaluation of ten brands of pre-poured Mueller-Hinton agar plates for EUCAST disc diffusion testing. *Clinical Microbiology and Infection* **28**, 1499.e1-1499.e5 (2022).

59. Lippmann, T. *et al.* Fluorescence *in Situ* Hybridization (FISH) for the Diagnosis of Periprosthetic Joint Infection in Formalin-Fixed Paraffin-Embedded Surgical Tissues. *The Journal of Bone and Joint Surgery* **101**, e5 (2019).

60. Jansen, G. J. *et al.* Rapid Identification of Bacteria in Blood Cultures by Using Fluorescently Labeled Oligonucleotide Probes. *J Clin Microbiol* **38**, 814–817 (2000).

61. Bentley, R. W., Harland, N. M., Leigh, J. A. & Collins, M. D. A *Staphylococcus aureus*-specific oligonucleotide probe derived from 16S rRNA gene sequences. *Lett Appl Microbiol* **16**, 203–206 (1993).

## Acknowledgements

This work was supported by the National Natural Science Foundation of China (NSFC) for the Excellent Young Scientists Fund (Overseas, 0214530013), NSFC for Distinguished Young Scholars (82302837), the China Aerospace Science and Technology Corporation (0231530004), the Strategic Partnership Research Funding (HUST-Queen Mary University of London, No.2022-HUST-QMUL-SPRF-07). We thank the Medical Subcentre of Huazhong University of Science and Technology (HUST) Analytical & Testing Centre. This project was approved by the China Spallation Neutron Source (CSNS) under the grant number P0123122900034 and we thank the staff members of the Small Angle Neutron Scattering (https://cstr.cn/31113.02.CSNS.SANS) at the CSNS (https://cstr.cn/31113.02.CSNS), for providing technical support and assistance in data collection.

## Author contributions

Y.W., N.G., Z.Y. and C.Z. conceived the project. Z.Y. and C.Z. carried out the experiments. Y.W. and N.G. supervised the study. L.X performed WF extracting. S.S performed biological characterization. D.Y conducted the SANS and analysed the data. S.Z conducted biofilm culture. Y.K. and H.Y. assisted with SANS. G.Z and J.S assisted with animal experiment.

**Competing interests**

The authors declare no competing interests.

**Correspondence and requests for materials**

should be addressed to Yuanhao Wu and Nengqiang Guo.

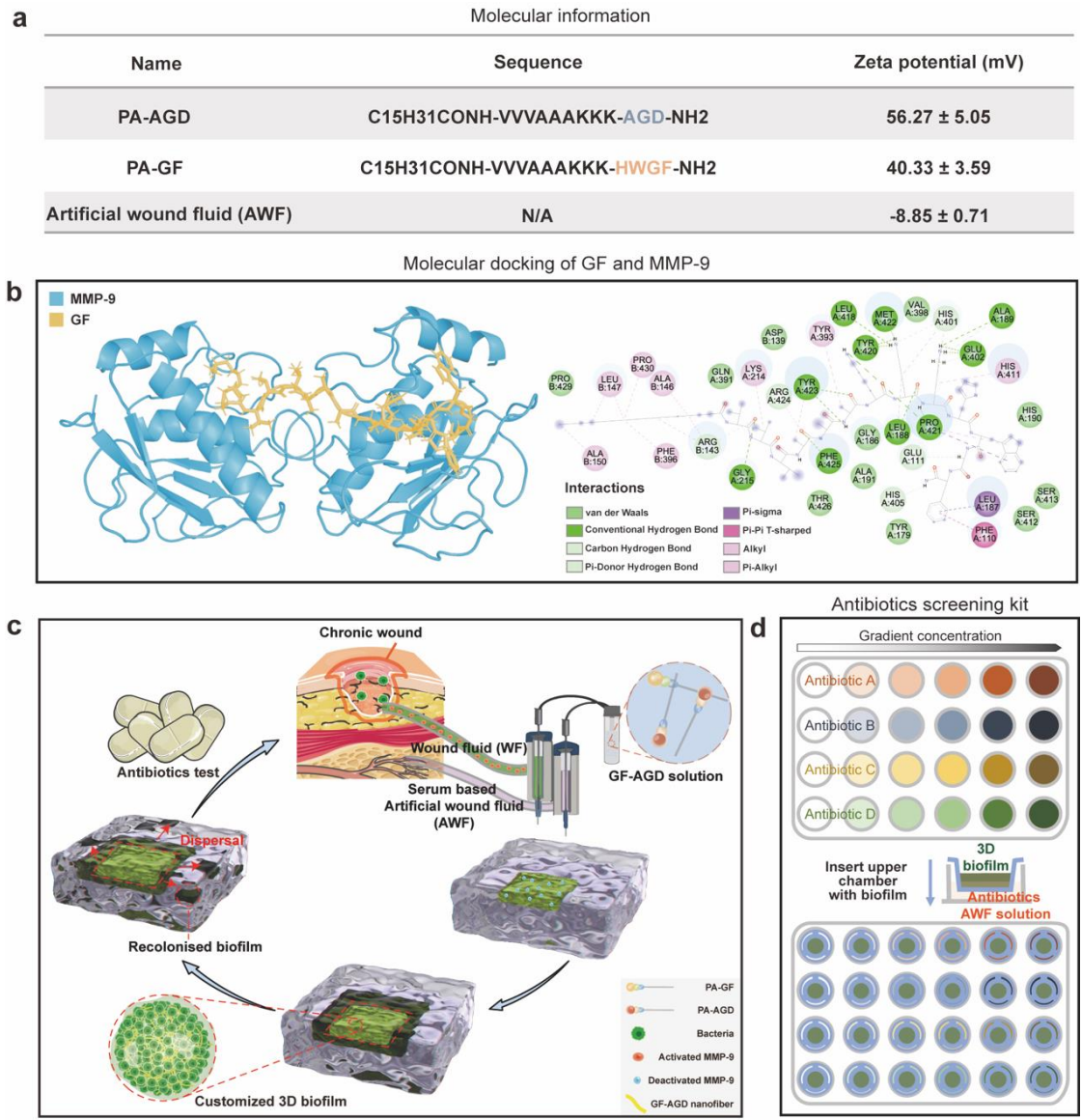

**Fig. 1 | Rationale for co-assembling system and antibiotics screening model. a**, Summary table of the key characteristics of the peptide amphiphiles (PAs) and artificial wound fluid (AWF). **b**, Molecular docking analysis showing that PA-GF binds tightly to the catalytic active site of one subunit within the MMP-9 dimer. **c**, Schematic illustration of the GF-AGD/AWF-based recolonisation biofilm model construction. **d**, Antibiotic screening kit established using Transwell inserts in a 24-well plate.

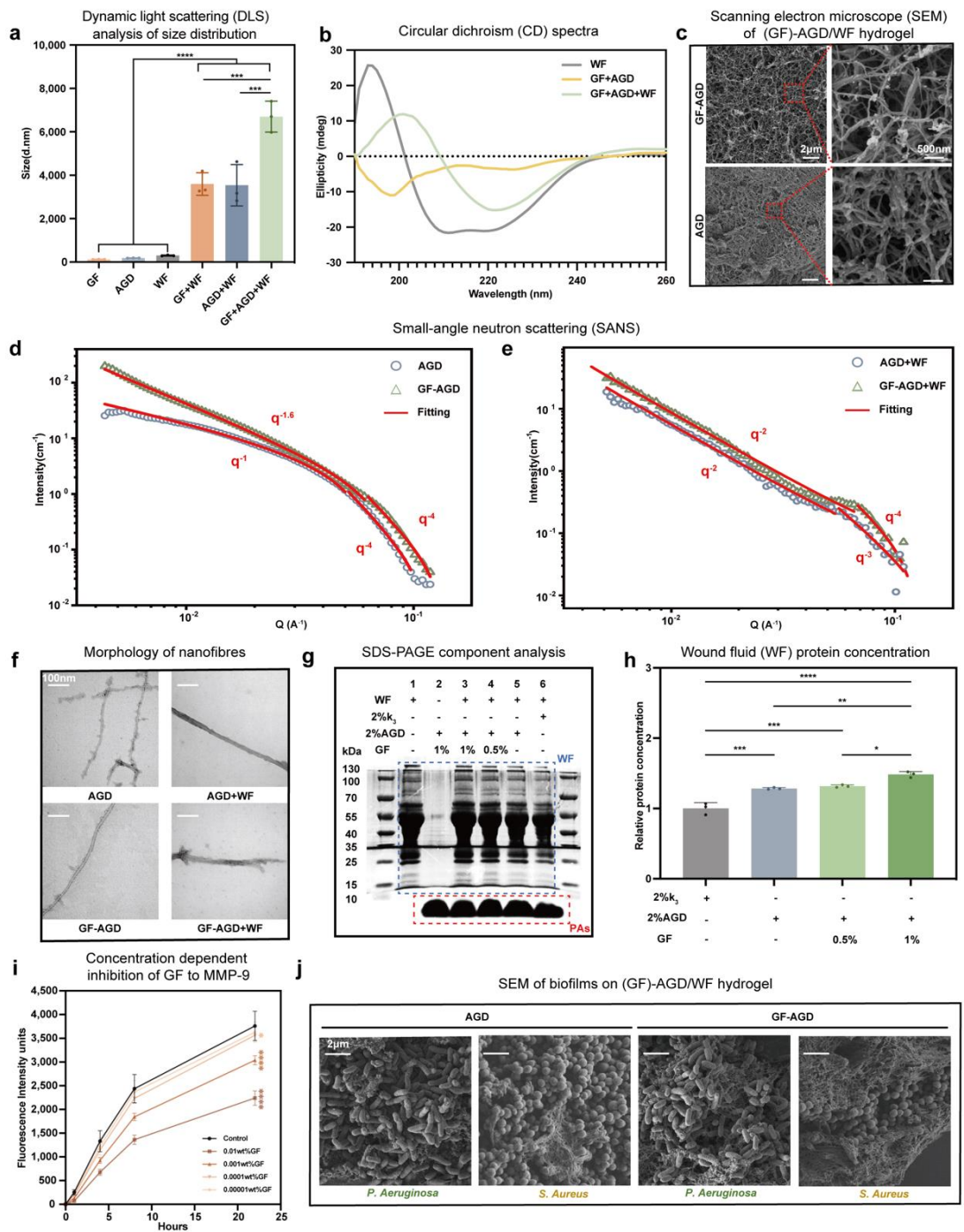

**Fig. 2 | Supramolecular Characterization and functional evaluation of PA/WF co-assemblies. a**, Dynamic light scattering (DLS) analysis showing particle-size distributions of PA-AGD, PA-GF, and their co-assemblies with wound fluid (WF). Data are presented as mean ± s.d. (n = 3). **b**, Circular dichroism (CD) spectra demonstrating a conformational transition from α-helix to β-sheet upon co-assembly of GF-AGD with WF. **c**, Scanning electron microscopy (SEM) images displaying homogeneous and fibrous nanostructure formed by PA/WF co-assemblies. **d**, Small-angle neutron scattering (SANS) patterns of 232-Å radius PA-AGD self-assemblies (blue circles) and 302-Å radius PA-GF (green triangles) rigid rod-like

structures. **e**, Co-assemblies with WF induced distinct scattering behaviours: PA-AGD/WF exhibited a $q^{-1.7}$ dependence corresponding to semi-flexible architectures of 450-Å radius, while GF-AGD/WF displayed a $q^{-2}$ dependence characteristic of branched morphologies of 350-Å radius. **f**, Transmission electron microscopy (TEM) images corroborating the formation of fibrillar assemblies consistent with SANS profiles. **g**, Sodium dodecyl sulphate-polyacrylamide gel electrophoresis (SDS–PAGE) analysis confirming efficient incorporation of WF-derived proteins within PA/WF hydrogels. **h**, Densitometric quantification of SDS-PAGE band showing enhanced protein retention with increasing PA-GF concentration. Data are presented as mean ± s.d. (n = 3). **i**, Fluorometric EnzChek™ assay demonstrating dose-dependent inhibition of MMP-9 enzymatic activity by PA-GF. Data are presented as mean ± s.d. (n = 3). **j**, SEM images showing mature biofilms of *Pseudomonas aeruginosa* (*P. aeruginosa*, ATCC 27853) and *Staphylococcus aureus* (*S. aureus*, SH1000) integrated within (GF)-AGD/WF hydrogels. NS at $P > 0.05$; ****$P<0.0001$; ***$P<0.001$; **$P<0.01$; *$P<0.05$ by one-way AVONA and Tukey's post-hoc test.

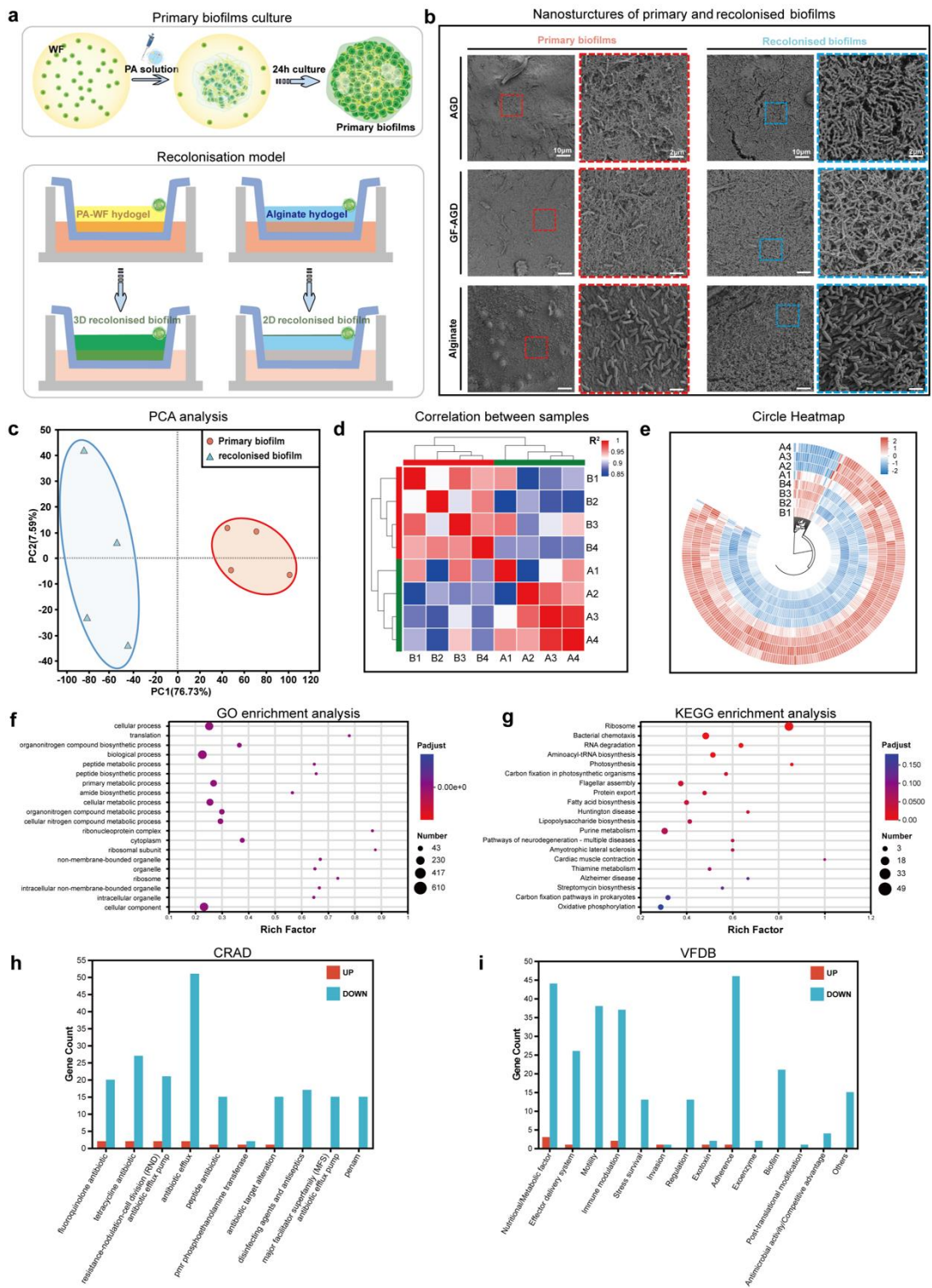

**Fig. 3 | Validation and transcriptional profiling of the recolonisation biofilm model. a**, Schematic of the recolonisation model established by connecting a fragment of mature PA/WF biofilm hydrogel to a pristine PA/AWF hydrogel layer within an insert cultured in 24-well plate. **b**, SEM images revealing morphological differences of primary and recolonised *P. aeruginosa* biofilms on alginate, AGD/WF, and GF-AGD/WF hydrogels. **c**, Principal component analysis

(PCA) showing distinct clustering between primary and recolonised *P. aeruginosa* biofilms, with PC1 and PC2 explaining 76.73 % and 27.59 % of variance, respectively. **d**, Pearson correlation heatmap demonstrating high intra-group reproducibility ($R^2 > 0.95$) and clear inter-group divergence. **e**, Hierarchical clustering heatmap of 1,138 differentially expressed genes (DEGs) between primary and recolonised biofilms. **f**, Gene Ontology (GO) enrichment analysis highlighting biological processes associated with cellular activity, translation, and organonitrogen compound biosynthesis. **g**, Kyoto Encyclopedia of Genes and Genomes (KEGG) analysis identifying enrichment in ribosome biogenesis, bacterial chemotaxis, RNA degradation, and flagellar assembly pathways. **h**, Comprehensive Antibiotic Resistance Database (CARD) annotation showing downregulation of multidrug efflux RND family genes in recolonised biofilms. **i**, Virulence Factors Database (VFDB) annotation revealing suppression of genes linked to adhesion, secretion systems, and metabolic virulence.

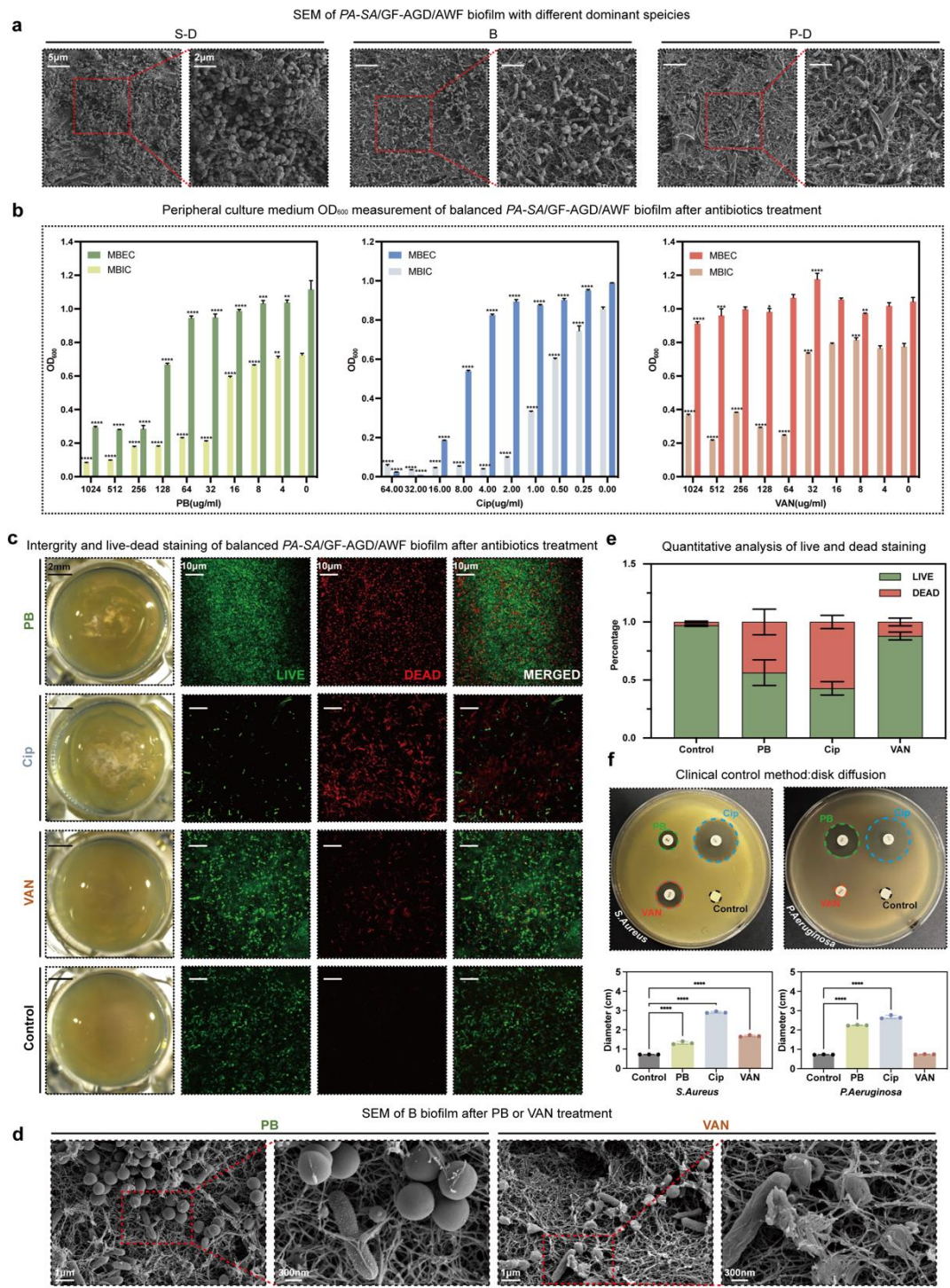

**Fig. 4 | Antibiotic screening kit based on GF-AGD/AWF polymicrobial biofilm. a**, SEM showing formation of *P. aeruginosa–S. aureus* dual-species biofilms within the GF-AGD/AWF hydrogel at different inoculation ratios, generating *S. aureus*-dominant (S-D), balanced (B), and *P. aeruginosa*-dominant (P-D) communities. **b**, $OD_{600}$ measurements of B-biofilms evaluating bacterial growth and regrowth after 24 h polymyxin B (PB), ciprofloxacin (Cip), or vancomycin (VAN) exposure. Data are mean ± s.d. (n = 3). **c**, Representative stereomicroscopic and live/dead staining images of post-treatment B-biofilms, illustrating that Cip completely

disrupted the biofilm macrostructure, PB partially disrupted the central region, while VAN had minimal effect. **d**, SEM images showing distinct ultrastructural damage of bacterial cell membranes after treatment with PB or VAN. **e**, Live/dead staining quantitative analysis confirming the superior bactericidal activity of Cip, followed by PB, with VAN exhibiting the lowest efficacy against dual-species biofilms. Data are mean ± s.d. (n = 3). **f**, Standard agar diffusion assays demonstrating consistent inhibition patterns with the *in vitro* GF-AGD/AWF screening platform. Data are mean ± s.d. (n = 3). NS at $P > 0.05$; ****$P<0.0001$; ***$P<0.001$; **$P<0.01$ by one-way AVONA and Tukey's post-hoc test.

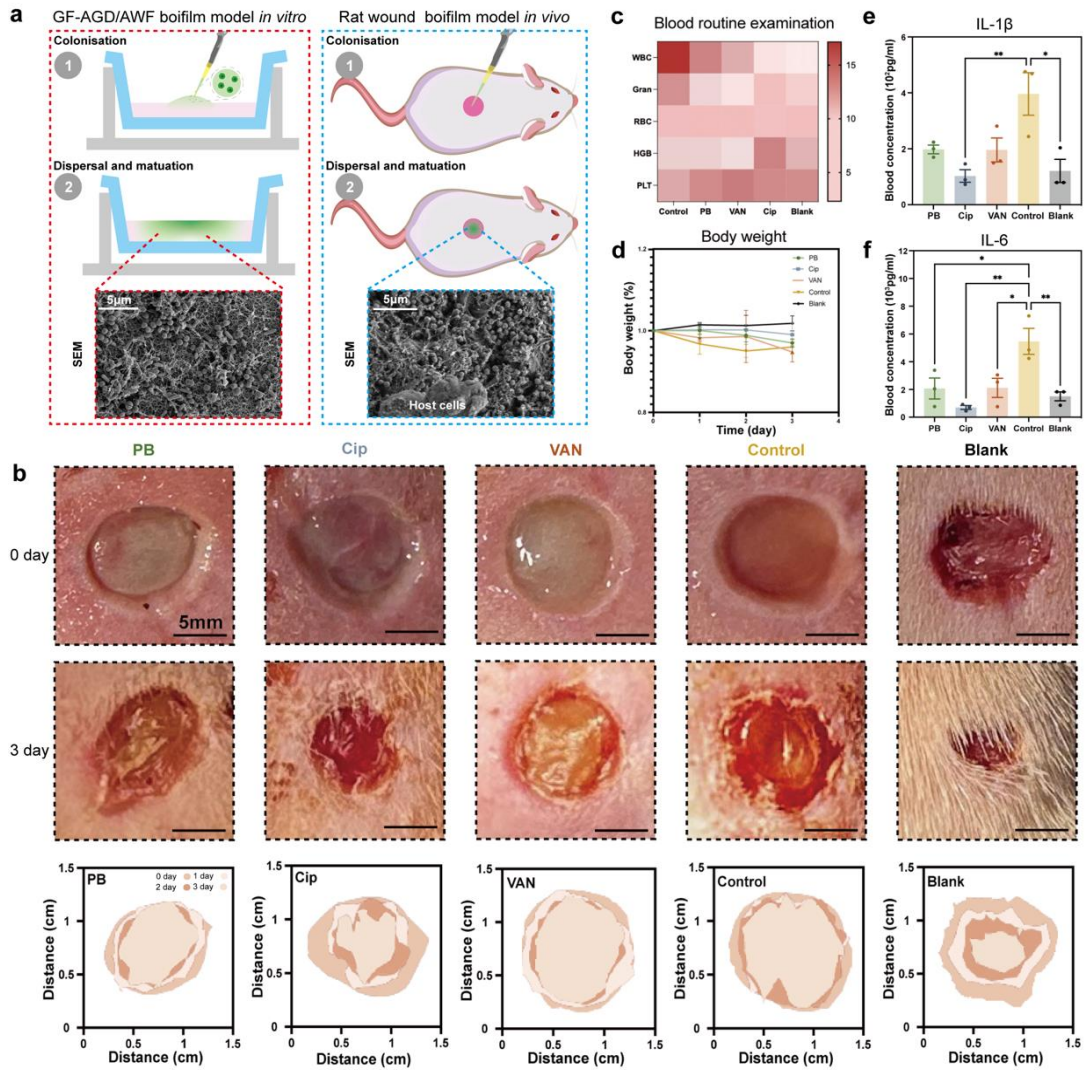

**Fig. 5 | *In vivo* validation of the GF-AGD/AWF antibiotic screening model using a polymicrobial chronic wound infection. a**, Similarity in biofilm formation process and microstructure between *in vivo* rat wound biofilm infections and the PA/AWF *in vitro* biofilm model. **b**, Establishing a full-thickness excisional chronic wound model in rats co-inoculated with *P. aeruginosa* and *S. aureus* (1:1), showing accelerated wound closure following Cip treatment compared with PB, VAN, and untreated control after 48 h of biofilm maturation and antibiotic exposure. **c**, Haematological profiling of red blood cell counts, haemoglobin concentrations, and platelet levels demonstrating no overt drug-related toxicity across treatment groups (unit: WBC-$10^9$/L, Gran-$10^9$/L, RBC-$10^{12}$/L, HGB-$10^1$/L, PLT-$10^{11}$/L), Data are mean (n = 3). **d**, Monitoring body weight fluctuations within ±10% throughout the experimental period, confirming overall animal health and procedural compliance, Data are mean ± s.d. (n = 3). **e–f**, Quantifying systemic in flammatory responses by measuring leukocyte counts and plasma concentrations of IL-1β and IL-6, showing lowest inflammatory levels in Cip-treated animals. Data are mean ± s.d. (n = 3). NS at P >0.05; **P<0.01; *P<0.05 by one-way AVONA and Tukey's post-hoc test.

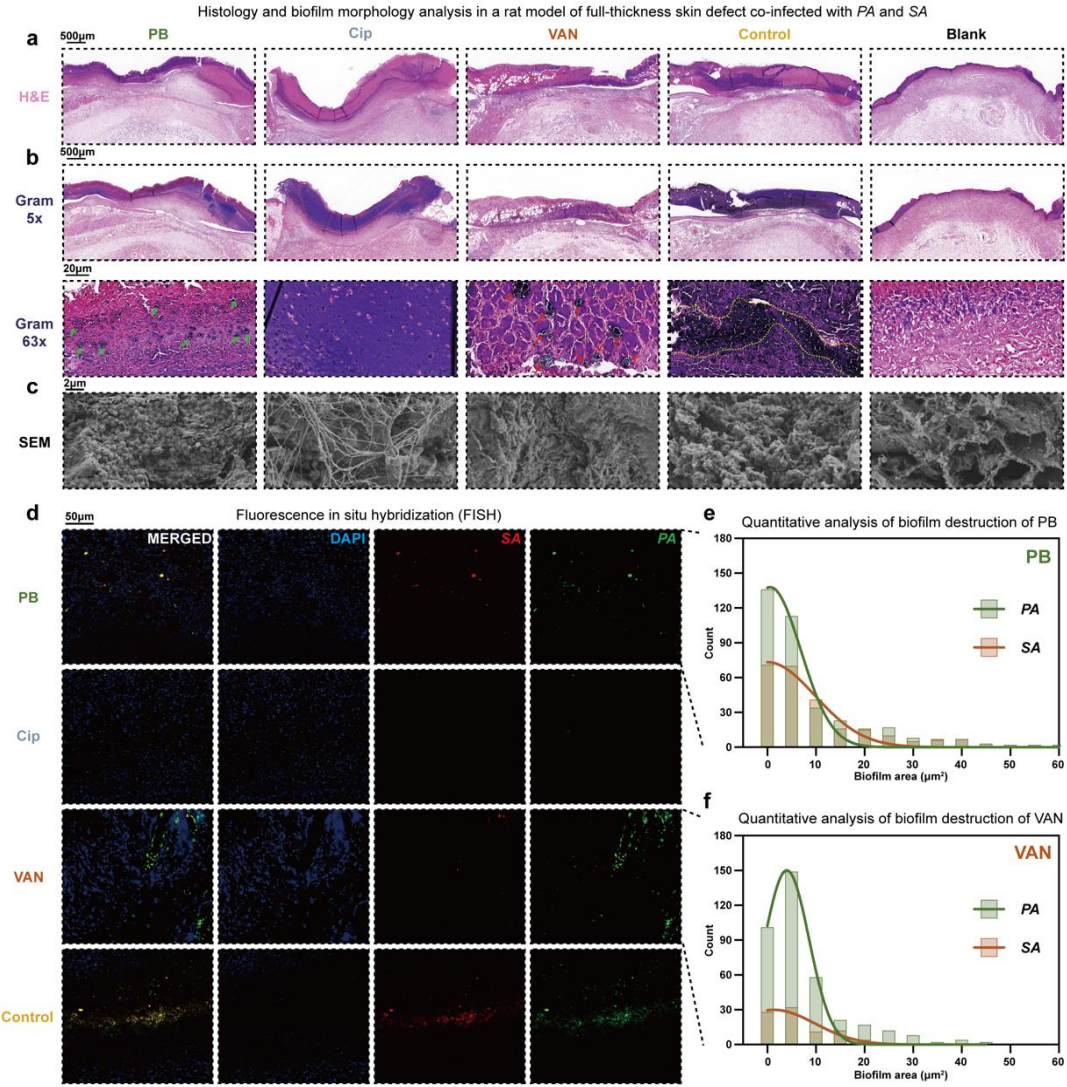

**Fig. 6 | Structural and microbiological validation of *in vivo* polymicrobial infection and its consistency with the GF-AGD/AWF *in vitro* model. a**, Haematoxylin and Eosin (H&E) staining of wound sections showing thick scab layers in all infected groups, indicative of microbial colonization. Cip treatment markedly reduced inflammatory cell infiltration compared with PB and VAN groups, validating the *in vitro* antibiotic screening outcomes. **b**, Gram staining of wound tissues revealing dense bacterial aggregates in untreated controls, smaller aggregates in PB-treated wounds (~5 μm, green arrows), and larger clusters in VAN-treated wounds (~10 μm, red arrows), while no visible aggregates were detected in the Cip group. **c**, SEM imaging of wound biofilms. **d**, Fluorescence *in situ* hybridisation (FISH) images showing co-localisation of *P. aeruginosa* and *S. aureus* within untreated wounds, with Cip eliminating both bacterial populations and narrow-spectrum antibiotics (VAN and PB) leading to altered microbial distribution. **e–f**, Quantification of bacterial biofilm aggregate sizes demonstrating the smaller biofilm aggregates following PB treatment than VAN-treated wounds.